\documentclass[runningheads]{llncs}
\usepackage[T1]{fontenc}
\usepackage{graphicx}
\usepackage{booktabs}
\usepackage[misc]{ifsym}
\usepackage[numbers]{natbib}

\usepackage{mwe}
\usepackage{hyperref}
\usepackage[outercaption]{sidecap}

\newcommand{\companyname}{}

\begin{document}

\title{Archetypes or ability? Clustering for modelling student mathematical competence}

\titlerunning{Archetypes or Ability? }

\author{Benjamin Mawdsley\inst{1}\thanks{Authors contributed equally} \and
Tom Quilter\inst{2}\thanks{Authors contributed equally} \and
Richard Turner\inst{3} \and
Sarah Jackson\inst{1} \and
Paul Edwards\inst{1}}

\authorrunning{B. Mawdsley, T. Quilter et al.}

\institute{STFC Hartree Centre \and
University of Manchester\\
\email{thomas.quilter@manchester.ac.uk} \and
University of Cambridge}

\maketitle              

\begin{abstract}
Personalised learning systems often assume that mathematical ability is combined of discrete abilities, acquired sequentially and dependent upon first acquiring foundational abilities, and students often report different strengths. In this work, we explore the validity of these assumptions by applying clustering methods to a large dataset of 119,034 students, spanning 13 national-level exams sat in the United Kingdom and collected by the \companyname platform. Classifying question results as pass or fail, we use a Bernoulli Mixture Model to search for latent populations which would be indicative of discrete skill-sets.  We find that few distinct clusters are present in the data and that the dominant factor is overall student ability, which is further supported by the high degree of linear correlation between the probability distributions of the resulting clusters. Our best performing model achieves an accuracy of 78\%, competitive with more complicated models in the literature whilst being more explainable.  Comparing this model's performance with logistic regression baselines and with k-nearest neighbours, we find a small improvement when using performance on each individual question as features. This suggests that whilst overall ability level is the dominant factor for predicting performance, small further personalisation improvements can be made by tailoring to a students exact strengths, but that students do not appear to develop strongly differing ability across topics. Our work offers a national scale test of machine learning in education and offers a new benchmark for the field, demonstrating how explainable models can reach competitive performance.  

\end{abstract}

\section{Introduction}

In the United Kingdom educational system, GCSE certificates are a significant qualification that define future opportunities. Students typically sit them aged 16, with passing grades in subjects like mathematics and English being frequent pre-requisites for further study or for employment. For the educator, cohort performance is a key statistic for monitoring educational performance \cite{OfstedGuidance}; the fraction of students leaving with 5 passing grades is a common metric used by parental guardians when choosing schools \cite{SchoolRanking}. Students and teachers therefore have a strong motivation to achieve positive results in these exams, leading to increasing pressure on all parties. \par 

Alongside teaching, preparation for GCSE examinations often includes sitting previous exams or ``mock" papers. Mock papers prepare students for the examination environment and to practice on representative questions. Crucially, mock exams serve as a way to estimate the current ability levels of the students and to identify potential areas of improvement. By combining this knowledge with their knowledge of in-classroom ability, the teacher can estimate an ``expected grade" for the student. Knowledge of what grades they are likely to achieve enables the student to plan their future choices and to set their expectations. Estimating predicted grades relies on an understanding of how the student is likely to develop in ability, the mathematical skills they have already mastered, and the remaining time for further study. \par 

Whilst most students are able to sit real GCSE examinations and receive a grade according to their performance, some students miss the opportunity and are instead required to use their predicted grades. For example, the Covid-19 pandemic brought this to public attention when a large scale algorithmic approach was used to assign grades, controversially introducing a wide disparity in student outcomes \cite{exams_fiasco,covid19-2https://doi.org/10.1002/berj.3705}. The subjective aspect of predicted grades can make them susceptible to biases, with the potential for specific groups to be systematically disadvantaged \cite{PredGRadDiscrim_doi:10.1080/09645292.2020.1761945}. As predicted grades can significantly impact the lives of students, a fair and accurate approach to quantifying their ability is crucial for the GCSE system. \par 

The modular format of mathematics curricula, and that students self-report confidence in different topics, suggest that a single grade may not capture all about a student's ability. Modelling discrete skill sets may offer a predictive improvement and help inform the structure of teaching courses. Mock exam scores collected by teachers allow intervention before the final exam, and the importance of understanding student ability has led to the development of personalised education platforms using this data. There is an expanding literature of applying data analysis approaches in education and whilst the industry is still developing, it offers significant promise for improving student outcomes \cite{ALDOWAH201913,ciolacu2017education,mukul2023digital,rastrollo2020analyzing}.  \par 

\subsection{Main Contributions}
Within education, clustering based approaches are less frequently used than supervised predictive algorithms, despite their use being motivated by the possible existence of differing competency profiles in the student cohort. Our work contributes to this field by performing one of the largest studies of clustering for understanding student ability in mathematics. We hypothesise that discrete ability profiles exist in the student cohort, characterised by different academic strengths and weaknesses. We use an explainable Bernoulli Mixture Model that produces clearly defined clusters that map directly to question competencies. By probabilistically assigning to clusters,  we enable uncertainty to be incorporated into later decision making and confirm these are well calibrated estimates. We show that a single factor ability model without different competencies is sufficient to reach near-best performance, presenting a clear quantification of the value of knowing question-specific competencies. Through showing clusters are often a linear rescaling of each other, we rigorously explore the idea that ability is the dominant factor in prediction and highlight some subgroups that deviate from this model.  \par 
 Our key novel contributions are therefore: 1) a new understanding of clustering model performance in modelling student attainment in mathematics using a large dataset; 2) a benchmark comparison with logistic regression models to quantify the importance of by-topic ability; 3) a demonstration of how clusters can be used to find unusual student profiles. \par 
 The data used in this paper is proprietary and belongs to the \companyname platform, but is planned to be released at a future date. \par

\section{Related Work}

The expansion of educational data has driven the growth of machine learning for understanding student development. Several meetings of the Conference on Neural Information Processing Systems (NeurIPS) conference have had dedicated ``education" challenges \cite{wang2020diagnostic,neurips2022}. In the 2022 competition, entrants sought to identify the causal connections involved in learning \cite{kumar2023conceptual}. \cite{neurips2020_1}  predicted which multiple choice answer would be chosen by students, and then whether that answer was correct, through a combination of feature extraction and ensemble models. Their scores of 67\% accuracy for predicting which answer a student would select, and 77\% for whether they would be correct, offers a useful baseline for a data-led understanding of student ability. The highest performing model for suggesting which question to recommend to a student used a meta-learning approach and achieved a 72\% accuracy \cite{ghosh-bobcat}. Using a different dataset, \cite{skill-difficulty} modelled student performance according to an interaction between question difficulty and student ability, adopting an approach from Item Response Theory. By parameterising ability and difficulty they achieved an 80\% accuracy across 49521 students. Their key finding, that ability is the main driver of predictive accuracy, led them to argue that teachers should prioritise broader reasoning over discrete skills, but found some suggestions that improvements may come from adding additional, currently uncaptured, context. Supplementing exam results with extra data, such as video images of learners studying, has been shown to be useful for predicting performance \cite{lit_multimodelchango2021improving}.  \par 

 Curricula are a common part of teaching, where subjects are taught in a way that is built on modular and sequential components. Related to this, significant research has sought to answer whether there are clearly defined ``types" of students with similar characteristics. Such clusters could arise from students learning mathematics through common pathways, or reaching similar skill-sets \cite{usiskin1982van}. For example, \cite{saarela2016predicting} found that predictive power could be improved using student scores on preceding difficulty level questions. This leads to two questions: do students really learn skills in such a discrete fashion, and if so, can we create an optimal curriculum?  \par 
One way of searching for archetypal profiles would be through statistical clustering methods, where similar students would be near in the data space. In a recent review, \cite{rastrollo2020analyzing} found that the majority of research sought to predict performance at University level and that there was a shortage of effort analysing school age results. This was highlighted as a particular concern due to the importance of identifying struggling students at the earliest opportunity. They continue to say that unsupervised learning is less frequently used, proving noisy when used on real datasets and that improvements are required. \par \cite{lit_cluster1_doi:10.1080/15248372.2021.1939351} clustered a sample of $\approx$ 2000 children based on their self-confidence across a range of mathematical competencies, finding that higher performing students did have an advantage in specific topics and lending support to the idea of modular ability development. Another study by \cite{lit_cluster2_doi:10.1177/0022219419881632} used data from children in the early years of education, clustering into 3 broad classes of  mathematical ability and relating them to other traits of child development. \cite{lit_cluster3_kae13b} used a small dataset of 88 students to explore how pairwise clustering, based on the development of mathematical ability over time, could lead to personalised training plans and found six different ways that students developed in their data.  \par \cite{lit_cluster4_salles2020didactics} used results from mathematics exams in France, finding 4 clusters using a k-means approach but found that presence in each group was dominated by whether students passed or failed at a specific task. As K-means clustering has been used across several studies and found to be highly predictive of a student's future performance  \cite{liu2022data,mohd2023identification},  it  suggests that student similarities are a highly useful measure. As the simplest form of k-means treats distances in all directions as equal, it is less sensitive to differences in the relative importance of each feature - an issue addressed by \cite{chen2023research}. However, a large scale study of clustering that spans a representative sample of the student body is lacking.  \par



\section{Data}
``\companyname" \footnote{https://www.pinpointlearning.co.uk/} is a platform that helps teachers track the performance of their students. After students complete mock exams, their results are uploaded on a question-by-question basis by the teacher for the platform to display for analysis. Each unique student has an identifier, as does each different exam, that enabling comparison across different schools and classes. A sample of this data was provided to the authors of this paper, for over 119,034 student exams in the United Kingdom.  

\subsection{Data cleaning}
As the data stored in the \companyname dataset has been manually uploaded, this creates a potential source of error. The data was extracted from the platform for thirteen different mock exams, involving a total of 119,034 unique students. 
Metadata accompanying the exams allowed us to ensure that all scores sat within acceptable ranges for that question. Rows with a negative score for a question (4 instances) were removed, as were any cases with scores above full marks for a question (7.5\% of all rows). Removal was chosen instead of capping at 100\% to ensure confidence in the scores achieved in each specific question and not from later in the exam. Whilst 7.5\% is a significant fraction of our data to lose to cleaning, the size of our dataset makes this manageable. From the experience of the \companyname platform administrators, it is often the case that teachers would submit one score for all remaining questions in an exam - for example, if a student got 4 marks in the final 5 questions, they would put a 4 in for question 15 instead of breaking scores down by question. In order to completely remove this potential source of contamination from unrelated questions, we drop these rows. 

\subsection{Data exploration}

Each exam was sat by an average of 24137 students, after the data cleaning procedure had been applied. Each student sat their own combination of exams, typically sitting 2 or 3 out of the 13 - Appendix \ref{app:exam_desc} explores in further detail how these exams are distributed by students. \par
\begin{SCfigure}[1.0][h]
\centering
    \includegraphics[width=0.5\linewidth]{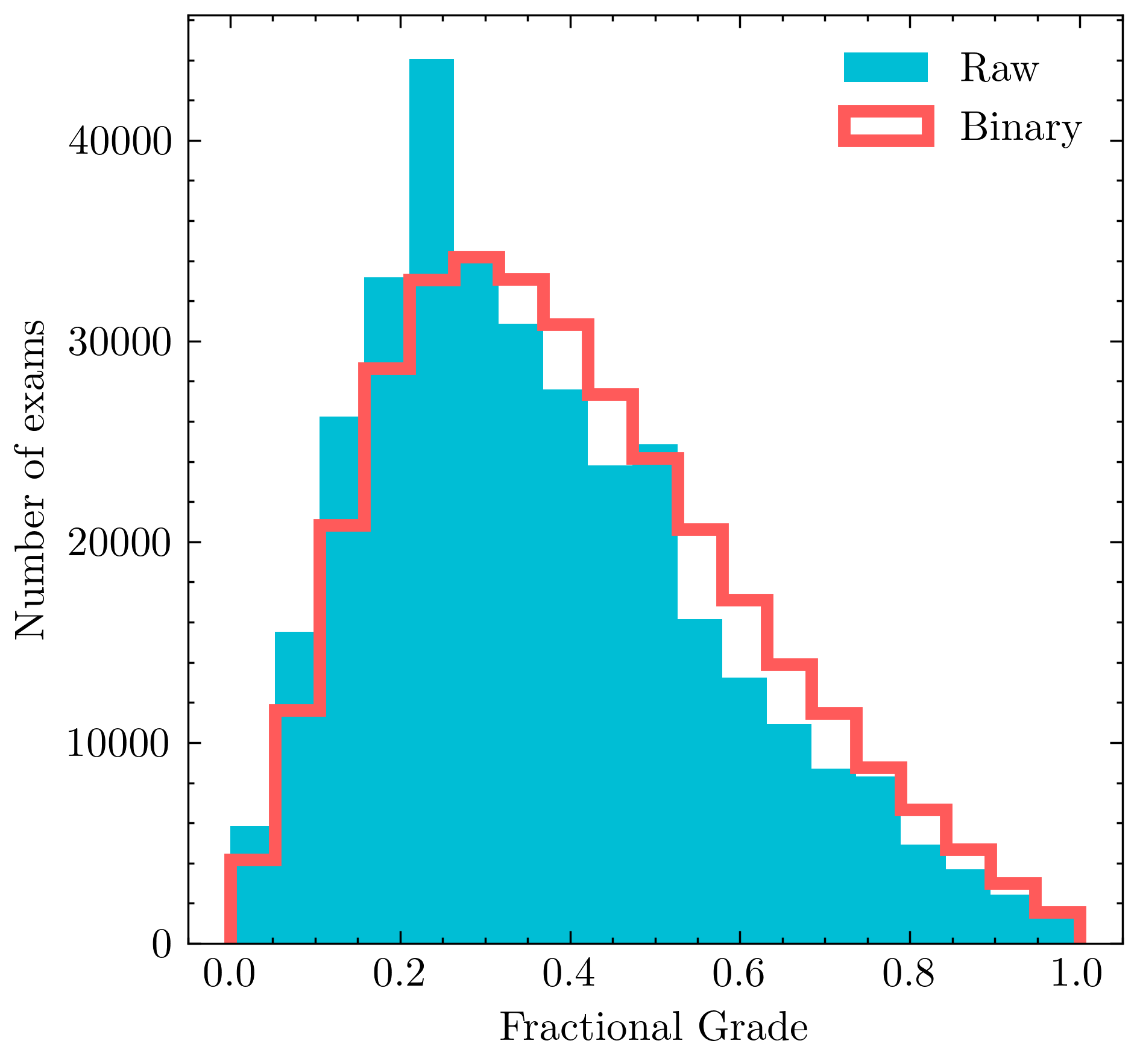}
    \caption{The distribution of overall grades for all completed examinations in our dataset. The solid bars show the original marks, and the step histogram shows the grades once each question has been converted to a binary pass or fail.}
    \label{fig:exam_score_dist}
\end{SCfigure}

 
Considering all exams, we have a broad representation of abilities in our dataset, as evidenced in Figure \ref{fig:exam_score_dist}. As these are \textit{mock} exams, taken for practice rather than to attain qualification, scores will be lower than those reported nationally.  A small number of students with very high grades can be seen in the tail of this distribution, with the bulk of grades around the 20\% to 40\% range and few scoring above 60\%. The dataset spans a representative range of abilities and and the distributions are of similar breadth in each exam.  \par 

The solid histogram in Figure \ref{fig:exam_score_dist} shows the exam scores calculated using the actual marks given on all questions, whereas the step histogram shows the overall score when it is calculated from questions which have been converted into binary pass/fail marks first. The distributions are very similar, with a slight shift to higher marks for binary marks; over half marks on a question is now equivalent to full marks, which can uplift some grades. \par

\begin{SCfigure}[1.0][!h]
    \centering
    \includegraphics[width=0.5\linewidth]{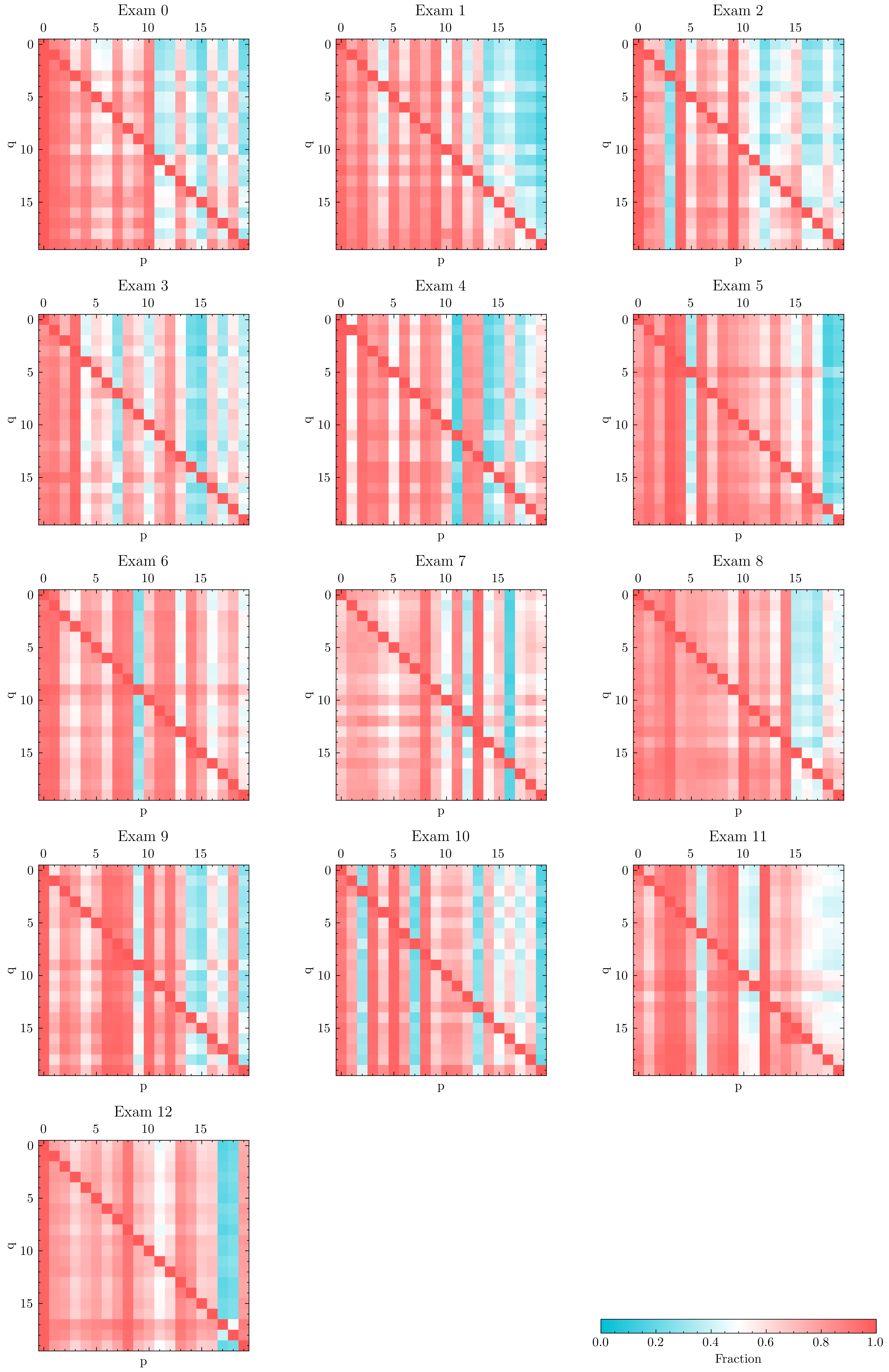}
    \caption{Exploring the predictive power of a passing grade in question $p$ when used to predict $q$, for all exams in the dataset. Colours describe the fraction of students who got a passing mark in question $p$ that also get a passing mark in question $q$. We only consider the first 20 questions as the final few questions are optional for the students. Top right parts of each matrix have a question $p$ earlier in the exam than question $q$, whereas the bottom left has question $p$ later in the exam than $q$. }
    \label{fig:qcorr_mat}
\end{SCfigure}

We explore how performance in one question changes the conditional probability that we assign to performance in another question in Figure \ref{fig:qcorr_mat}. This figure shows the average grade on a question $b$, given that a student has scored a passing grade on question $a$, i.e. $P(b=1|a=1)$. This matrix is asymmetrical as we calculate from different subsets each time (those that pass question a for $P(b=1|a=1)$, vs those that pass question b for $P(a=1|b=1)$). The upper right section shows later questions conditioned on earlier questions; probabilities in the upper right section are typically lower, as average grades are lower in the latter half of the exam but most students complete the early questions. Conversely, students with a passing grade in later questions are more likely to have also answered earlier questions successfully, and the lower left section of the matrix is more positive than the upper right. The different dependencies between question pairs are key inputs of the predictive model. \par  

\section{Modelling approaches}


In this section, we detail two clustering methods we use to model our examination data, alongside three versions of ''baseline" methods which do not assume the presence of clusters in our data. Each outputs a probability on the student getting a given question correct. \par

\subsection{Baseline models}

Our baseline models assume that student ability is a smoothly evolving function, with independence between the performance on the different questions. We therefore use logistic regression for these models, in two forms. We consider a logistic regression model:

\begin{equation}
    p(x) = \frac{1}{1+e^{-\beta_0 + \beta_iX}}
    \label{eq:logreg}
\end{equation}
where $p(x)$ is the probability of  $X$ is the input data, and $\beta$ are parameters learnt from the data. The $\beta$ features vary across the two models:\par

\begin{itemize}
    \item \textbf{Single ability model  (LogReg$_s$):} Models students according to an average grade across all questions, resulting in a single $\beta_i$ variable.
    \item \textbf{Subject specific dependencies (LogReg$_m$) :} The score on every other question on the exam is a feature, such that there are now $\beta_i$ contains 19 variables for the 20 question exam. This allows some questions to have a stronger influence on the prediction than some others.
\end{itemize}

In addition to these, we calculate  a \textbf{constant probability model (Flat)} which assumes a constant probability for all students, estimated from the general pass rate on a question estimated from the whole population used to train the model.


\subsection{Clustering models}
We use two varieties of clustering models to investigate whether closely related groups of skills exist in our data:
\begin{itemize}
    \item \textbf{$k$-nearest neighbours(kNN):} assigns probabilities based on the average score of the $k$ nearest points in the feature space. We use the \\\texttt{sklearn.neighbors.KNeighborsClassifier} from the scikit-learn package \cite{scikit-learn}, using a Euclidean distance measure. 
    \item \textbf{Bernoulli Mixture Model (BMM):}  Clusters are defined by their probabilities of getting each question correct, and students are probabilistically assigned to each cluster. Fitting these clusters requires an Expectation-Maximisation (EM) approach, and implement our own python code to achieve this using the E and M steps described in \cite{carreira2000practical}. Students have a probability of being a member of each cluster, which is combined with the probability of getting the question right in each cluster to give a final predicted probability for each student. 
\end{itemize}

\section{Experiments}
\subsection{Experimental design}
if clusters exist in our data, they should produce a predictive performance that is above our baseline models. Splitting by cohorts that sit each exam, we iterate across each exam, using each question as a target variable in turn and the remaining 19 questions as potential features. We split the data with a 90\% training dataset and a 10\% testing dataset, held out for final evaluation. Within that 90\% training dataset, we perform k-fold cross validation with 8 folds per question to explore the optimal parameters for our models.   Each student sits each exam a single time, meaning each student is either a test sample or a training sample and no leakage can happen. \par 
For the kNN and BMM approaches, we optimise their parameters to produce the best performance on the validation dataset before reporting performance on the final test dataset. For the BMM, we explore the number of clusters using a range defined by [2,4,8,9,10,11,12,13,14,15,16,18,20,24,28,32, 64]. For the kNN's number of neighbours used, we explore a parameter range encapsulating all odd values between 3 and 20, and 57 further values regularly spaced up to 400.

We use several key statistics for quantifying the performance of each classifier: \textit{Log loss} for quantifying the predictive power, \textit{Accuracy}, \textit{Recall}, \textit{F1} statistic to account for the unbalanced dataset, and the \textit{Matthew's Correlation Coefficient } (MCC) for measuring the level of agreement between the binary labelled classes.
Error bars are estimated taking these measures during every k-fold validation run for samples in the 'train' dataset, and once for every question when evaluated on the 'test' dataset. Whilst we also calculate the performance of a constant probability assumption, most plots omit this measure as it is significantly below that recorded by all models. \par

\subsection{Predictive performance}

\begin{table}[]
    \centering

\begin{tabular}{|c| c | c | c | c | c |}
\hline
Model & Accuracy & F1 & Log loss & MCC & Recall \\
\hline
\hline
LogReg$_m$  & 0.778$\pm$0.005 & 0.744$\pm$0.011 & 0.463$\pm$0.007 & 0.334$\pm$0.009 & 0.759$\pm$0.014 \\ LogReg$_s$   & 0.769$\pm$0.005 & 0.728$\pm$0.013 & 0.477$\pm$0.007 & 0.289$\pm$0.009 & 0.749$\pm$0.016 \\ KNN  & 0.775$\pm$0.005 & 0.747$\pm$0.011 & 0.471$\pm$0.007 & 0.321$\pm$0.009 & 0.769$\pm$0.014 \\ BMM  & 0.776$\pm$0.005 & 0.74$\pm$0.011 & 0.466$\pm$0.007 & 0.324$\pm$0.009 & 0.754$\pm$0.015 \\ Flat  & 0.514$\pm$0.015 & 0.428$\pm$0.023 & 1.013$\pm$0.12 & 0.0$\pm$0.0 & 0.604$\pm$0.03 \\  \hline \end{tabular}
    \caption{Performance metrics for each of the models in this analysis. Metrics are calculated for every test question in the data, across all exams, for a total of 260 samples. Error bars are calculated by taking the standard error on the mean. }
    \label{tab:model_performance}
\end{table}

\begin{figure*}[!ht]
    \centering
    \includegraphics[width=\linewidth]{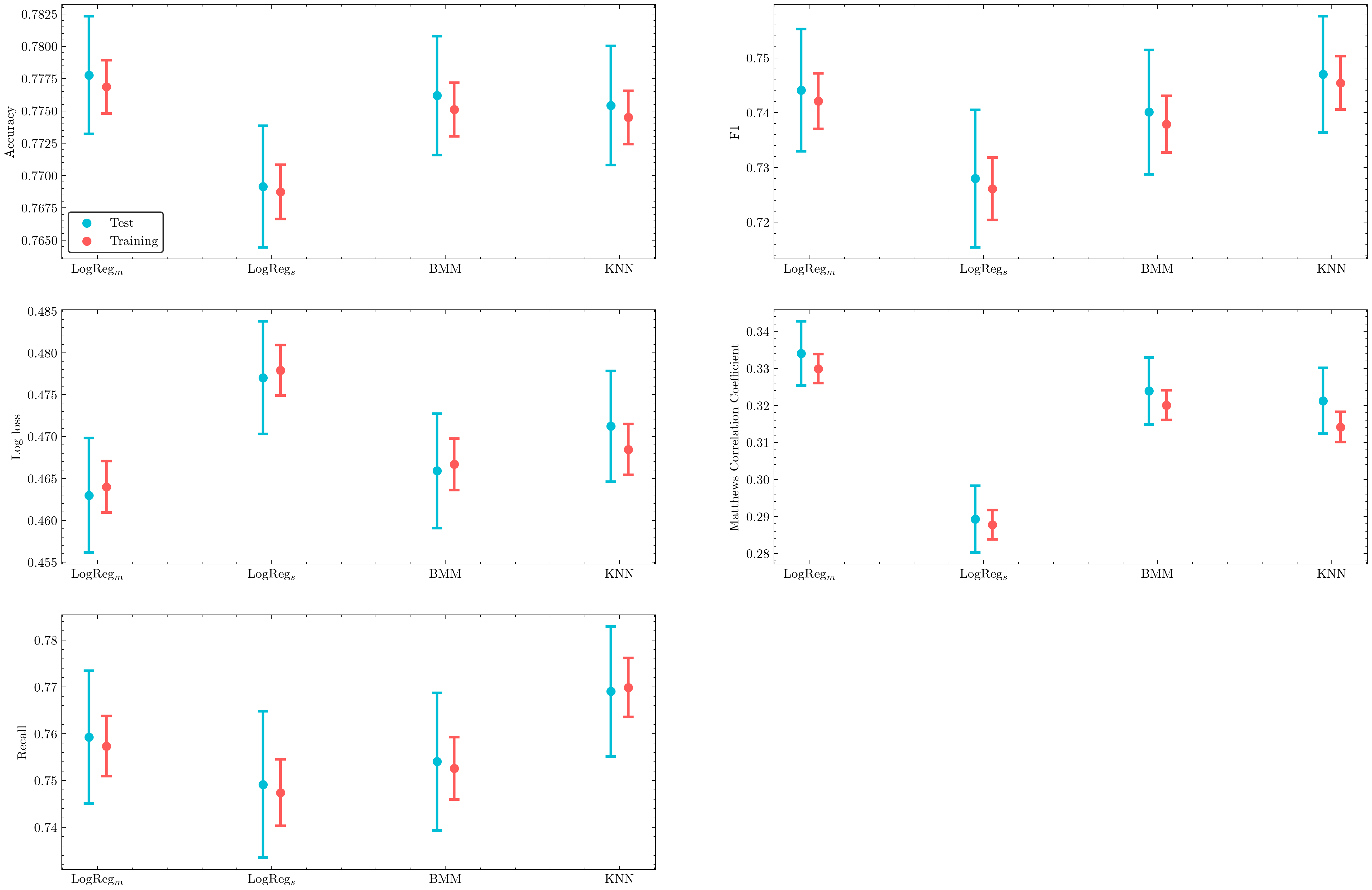}
\caption{Performance metrics for each of the models in this paper. Baseline performance, for a minimally informed classifier that assumed a constant probability for all students, performs significantly worse than all models shown.}
    \label{fig:pred_perf}
\end{figure*}

The performance of each model is shown in Figure \ref{fig:pred_perf} with the exact numbers shown in Table \ref{tab:model_performance}. In Figure \ref{fig:pred_perf}, the 'Training' dataset is the performance on validation sets through the k-fold cross validation process which are used for model selection, whereas the light blue 'Test' results are those from the final 10\% test dataset. \par 
Performance across all model classes is broadly similar across all metrics. The lowest log loss is for the multi-question logistic regression, shortly followed by the Bernoulli Mixture model, KNN and finally the single feature logistic regression. The width of the error bars show a large amount of overlap between these approaches. Despite adding many more features, the performance improvement between a single feature logistic regression and a multi question performance is quite small, supporting the conclusion from \cite{skill-difficulty} that the dominant factor in predicting performance is overall mathematical competence. 
For accuracy, the multi question logistic regression performs best, although all approaches have an accuracy of around 78\%. This is much higher than the baseline
model achieves (51.4\%, not shown in plots due to the large difference in scale with other model performance changes). All three of the BMM, KNN and multi question logistic regression score more highly than the single feature logistic regression, indicating that some secondary signal may be present beyond just student overall ability but that is a weak contribution. Interestingly, the KNN approach scores highest for recall, meaning that it captures more of the students who get a question correct. In cases where capturing which students know a topic is important, such as when prioritising which topics to revise, it may be the better suited model. This high recall score results in the KNN having the highest $F_1$ score despite its log loss being higher than both the BMM and multi-question logistic regression. For general performance and producing a well calibrated probabilistic estimate of whether a student is likely to know a given topic, the multi question linear regression performs strongest.   

\subsection{Probabilistic calibration}
\begin{SCfigure}[1.0][!ht]
    \centering
    \includegraphics[width=0.5\linewidth]{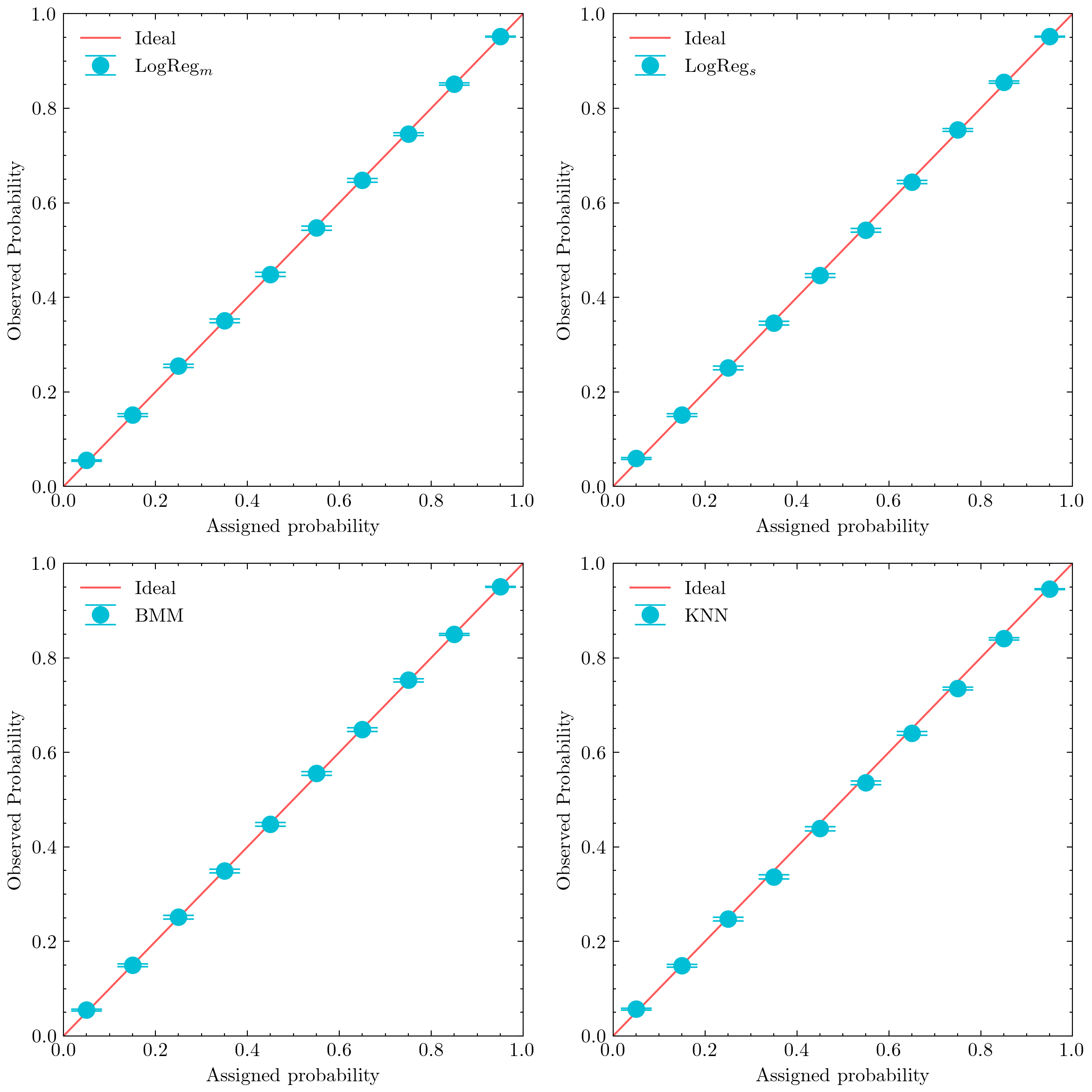}
    \caption{Ensuring that model probabilities are well calibrated. Each model outputs predictions for the questions in the test dataset and are binned. The rate at which students actually pass these questions is calculated within each bin and plotted on the y axis. The solid line indicates a perfect one-to-one correspondence between the assigned probability and the observed rate. All models have very well calibrated probability outputs, sitting close to the ideal line.}
    \label{fig:prob_calibration}
\end{SCfigure}

We explore how well the model output probabilities are calibrated by comparing to the truly observed probabilities, shown in Figure \ref{fig:prob_calibration}. The solid line shows a direct 1:1 correspondence, where probabilities assigned by the classifier would happen exactly as frequently as truly observed. Error bars are calculated using the binomial confidence interval function \texttt{binom\_confint} from the \texttt{scipy} package. Well calibrated performance can be seen, where all models lie very close to the ideal line, across all probability bins and for each model. 

\subsection{Cluster model properties}
Although the performance of the clustering approaches do not seem to yield significantly improved performance, we visualise these clusters to see what further insight their structure can give.  \par
\subsubsection{k-nearest neighbours}

\begin{figure}[!ht]
    \centering
    \includegraphics[width=\linewidth]{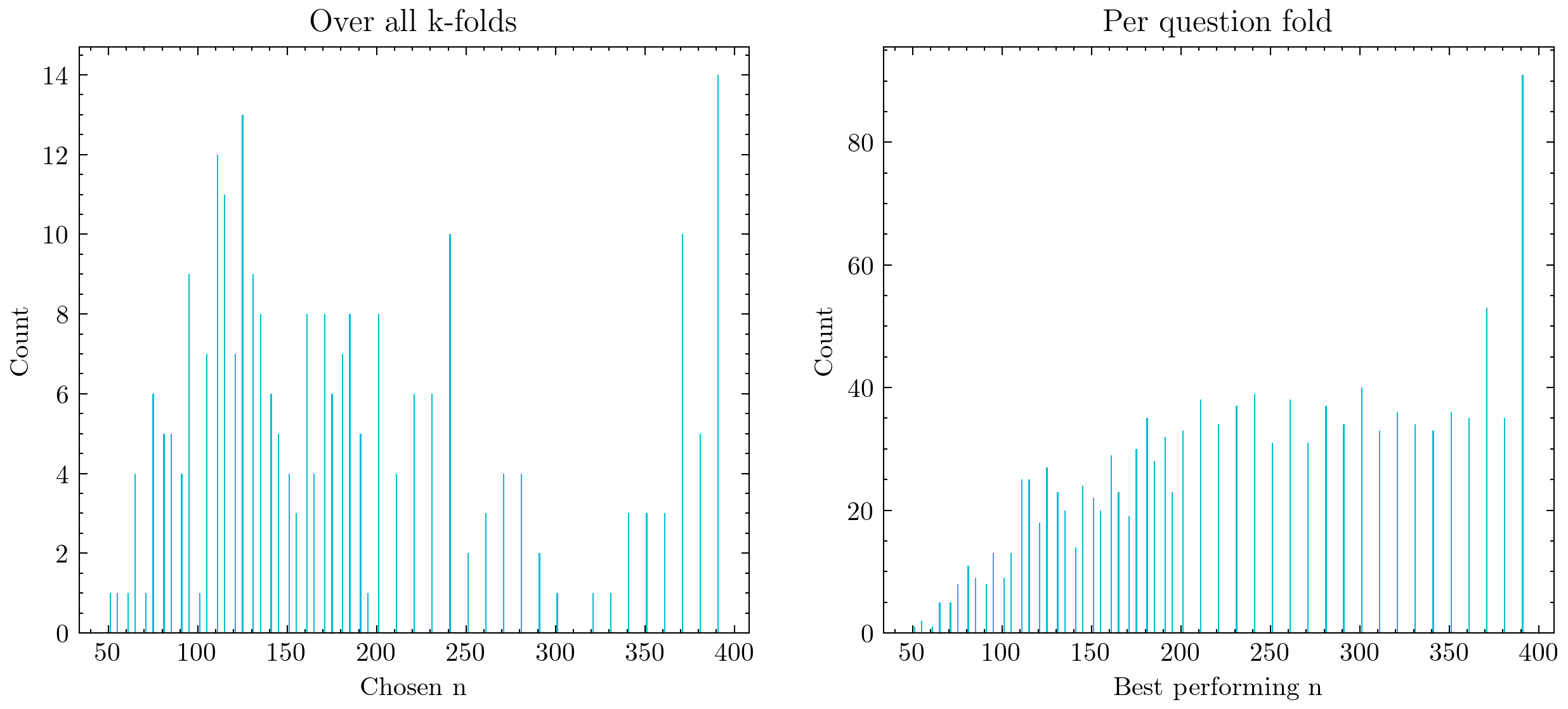}
    \caption{The selected number of nearest neighbours across each of the k-folds (left) and the best performing across all k-folds for a given question (right). The optimal number of nearest neighbours varies for each question, spanning the parameter range.}
    \label{fig:knn_chosen_k}
\end{figure}

The distribution of the number of nearest neighbours chosen for the kNN across all questions shown in Figure \ref{fig:knn_chosen_k}. When considering all folds, the best fitting number of clusters varies with peaks around $k=120$, but spans the parameter range explored. The best performing models have a stronger preference to higher values of $k$. Larger values of k allow the model to get a better estimate of the typical ability of students in the nearby area, whilst diluting the influence of the most similar students. This is similar to the story told by the linear regression models, where the dominant factor is overall ability, but these largest clusters still only represent approximately 2\% of the population.    \par

\subsubsection{Mixture model}

\begin{SCfigure}[1.0][!h]
    \centering
    \includegraphics[width=0.5\linewidth]{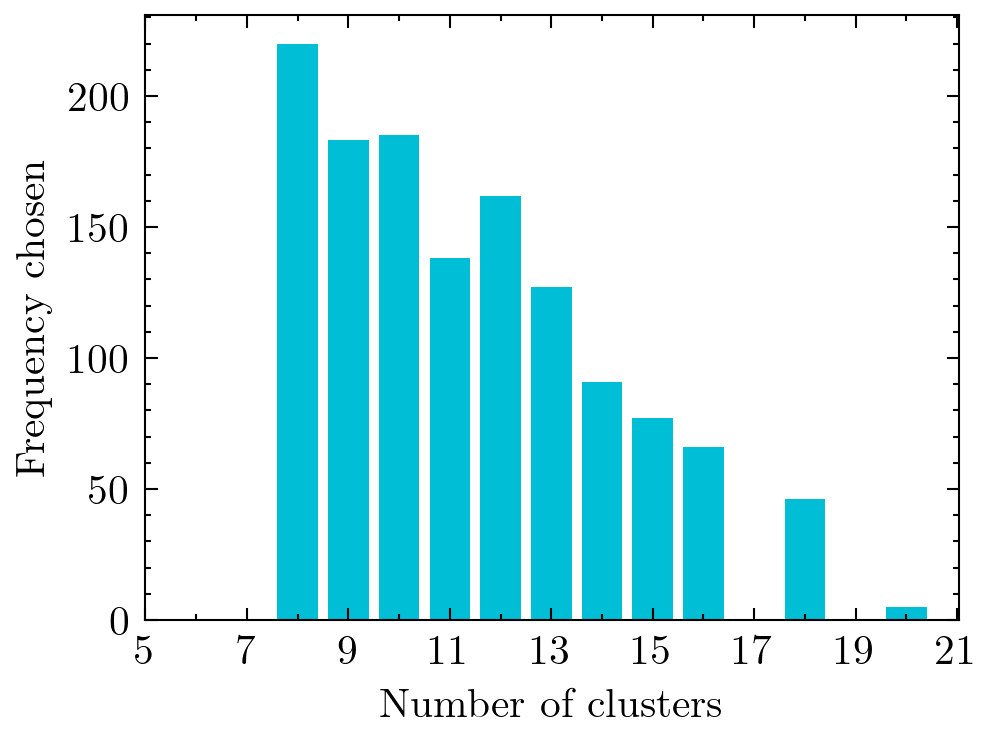}
    \caption{The optimal number of clusters chosen using the BMM approach, when taking the lowest BIC across the training folds for a given question in the dataset. Very few questions call for a high number of clusters, despite the parameter space explored extending to up to 64 potential clusters.  
    }
    \label{fig:bmm_chosen_dist}
\end{SCfigure}

The number of clusters chosen for the BMM approach is shown in Figure \ref{fig:bmm_chosen_dist}, where we choose to optimise the Bayesian Information Criterion (BIC) across the k training folds for a given question. This metric penalises more complex models if they do not produce a corresponding improvement in performance. We allow the number of clusters to be as high as 64. The fact that no runs of the fitting found these high values of clusters to be optimal means that different, specialised skill sets for students at each ability level do not generally seem to exist at any high number. Most questions select around 10-12 clusters, with a decaying frequency of optimisation runs finding more clusters.   \par

\begin{SCfigure}[1.0][!ht]
    \centering
    \includegraphics[width=0.5\linewidth]{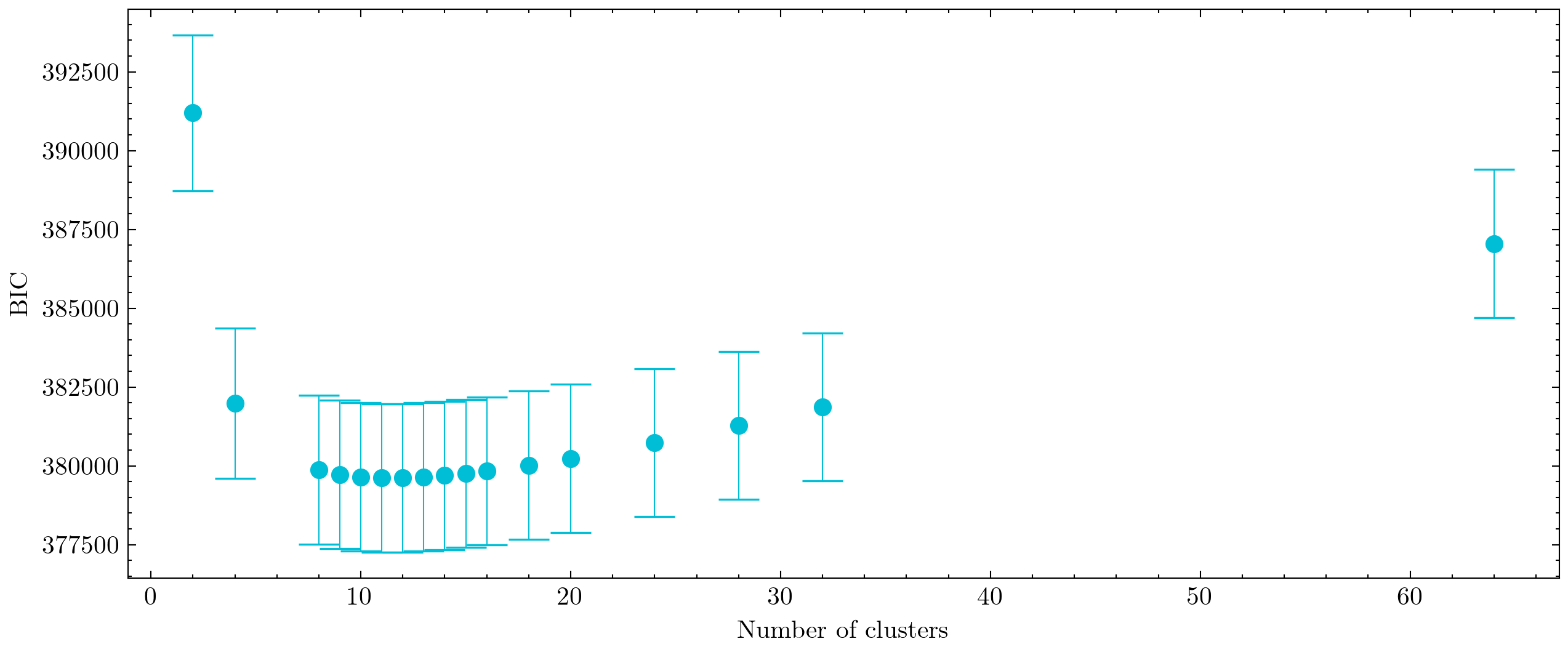}
    \caption{The distribution of BIC scores across all of the k-fold BMM models in our parameter exploration. A minimum can be seen around the 10-14 range corresponding to the most frequently chosen values in Figure \ref{fig:bmm_chosen_dist}.  
    }
    \label{fig:bmm_bics}
\end{SCfigure}

The distribution of BIC scores across all validation runs is shown in Figure \ref{fig:bmm_bics}. The mean value is plotted, with error bars being the standard deviation across all runs of the model fitting. Large variation for each of these numbers of clusters, when examined across all questions in the dataset, shows how different values can be optimal for different questions, such as shown in Figure \ref{fig:bmm_chosen_dist}. High numbers of clusters are penalised by the BIC metric if they do not provide sufficient performance improvements.

Whilst the number of clusters is informative for how varied the population appears to be, there is further insight we can take from the BMM output. The cluster coordinates output by the BMM define student archetypes in our dataset, with characteristic strengths and weaknesses, and therefore their exact shape could be illustrative. 
\par

\begin{figure*}[!ht]
    \centering
    \includegraphics[width=\linewidth]{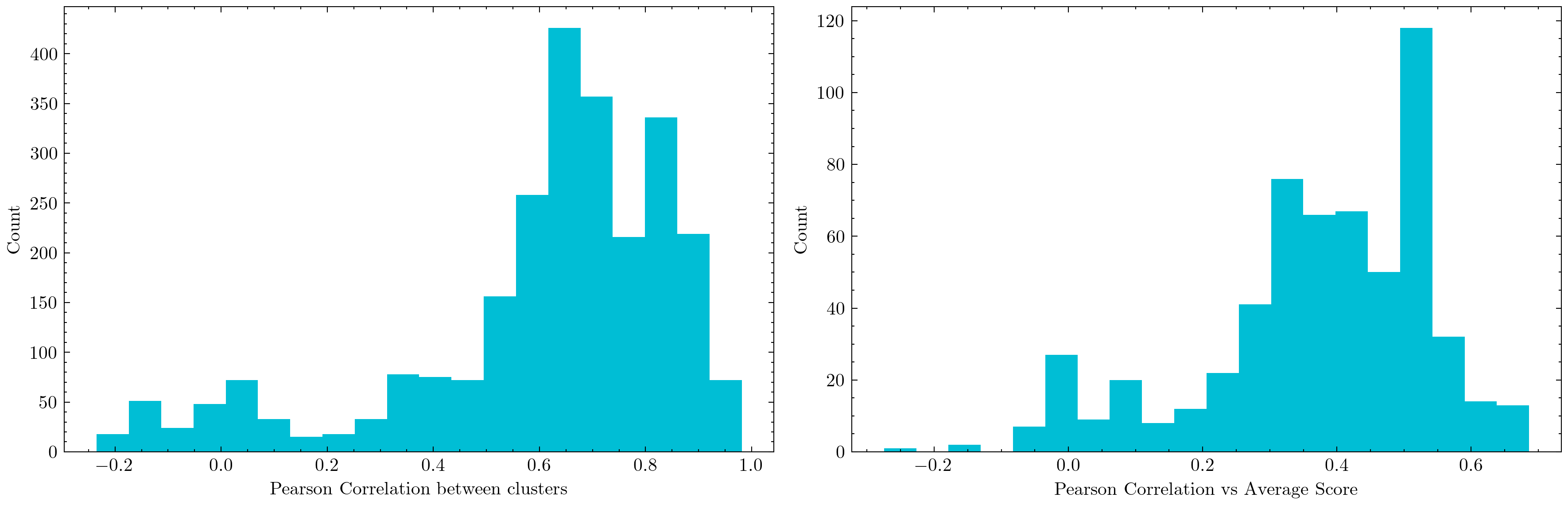}
    \caption{ The clusters have a very similar shape, dictated by the typical scores students get. Linear correlation between clusters means that the dominant difference between the clusters is a uniform amplitude change across each questions, not the shape of the clusters. A small number of clusters have a different shape to the overall exam scores (see Figure \ref{fig:bmm_leastcorr} for an example of a weakly correlated set of clusters.)}
    \label{fig:intercluster_linear}
\end{figure*}

We explore how much overall ability dominates the cluster shapes by linearly correlating clusters with each other; if performance improved uniformly across all questions as ability improved, then the cluster profiles would be highly linearly correlated, even if the exact pass rate at each question varies. We use the Pearson correlation coefficient to quantify this linear relationship, and measure this value between all pairs of clusters fitted to a given run of the BMM. Figure \ref{fig:intercluster_linear} shows the distribution of correlation scores across all of these cluster pairs. The left plot shows how the vast majority of clusters are strongly correlated with each other, with most clusters having a linear correlation above 0.6, corresponding to a strong linear correlation. The distribution in the right panel of Figure \ref{fig:intercluster_linear} shows each of the clusters correlated with the average score across all questions, with this correlation with the average performance showing that most clusters are a rescaling of average grade, but scaled according to different abilities. The dominant factor is ability and performance on each questions typically scales linearly with overall exam performance. A small number of cluster pairs have a near zero or slightly negative correlation, indicating that there may be small subsets of the population that have atypical behaviour and may benefit from increased personalisation. A further exploration of the profiles of these clusters is in Appendix \ref{app:intercluster_corr}. \par

\subsection{Consistency in model predictive power}
If this algorithm were to be used for suggesting appropriately challenging questions to real students, it is important to ensure that all students would be impacted similarly by the outputs of the model. For example, a model that had much better accuracy for male students vs female students would risk embedding biases into student development. A model therefore needs to be reliable for all students, and potential biases quantified. The data provided for this paper did not include demographic information, but we explore how performance varies across different levels of ability. This serves two purposes - it examines how ability is driving the predictions, and allows us to quantify how it would impact students of different abilities. \par

\begin{SCfigure}[1.0][!ht]
    \centering
    \includegraphics[width=0.5\linewidth]{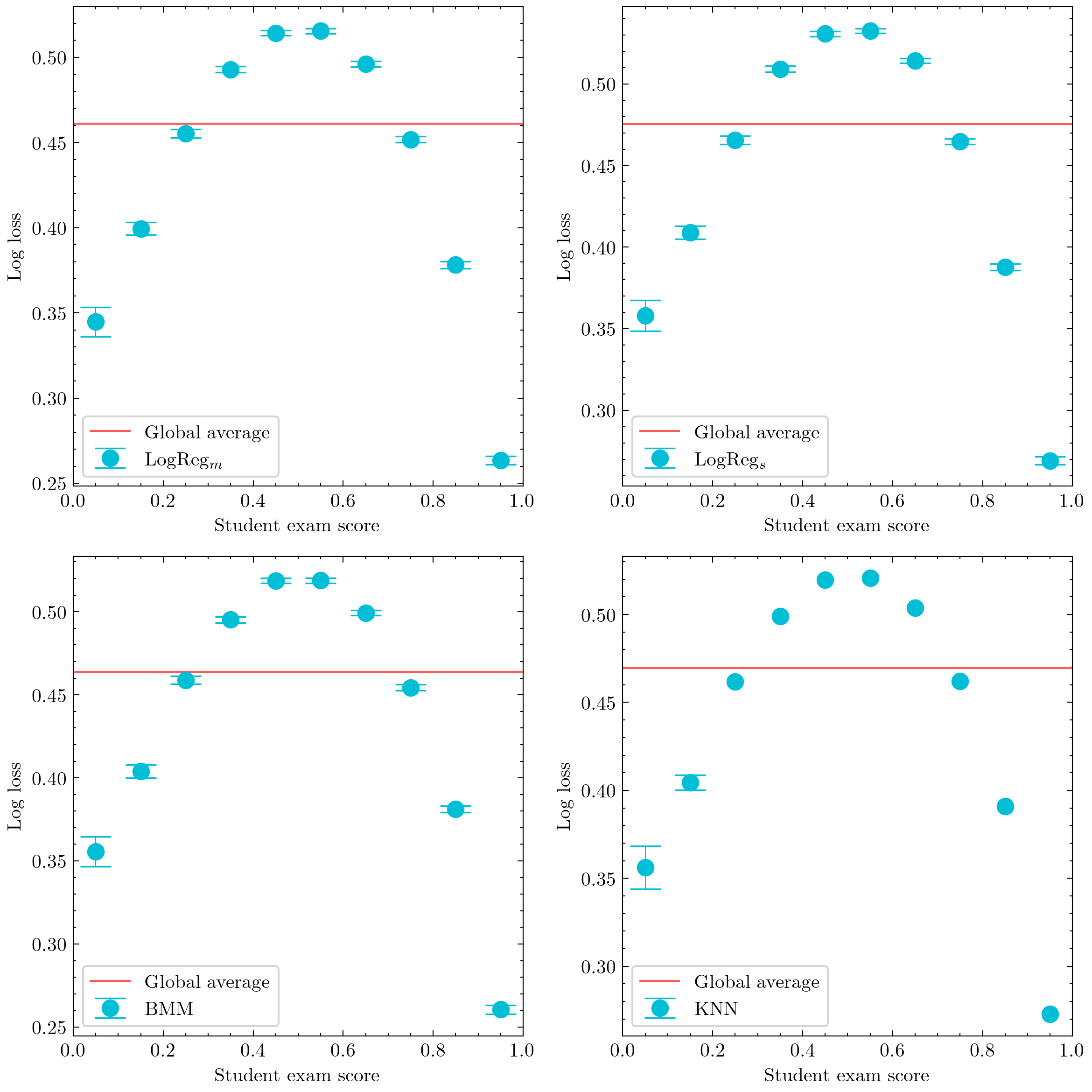}
    \caption{Model performance by ability cohort, where students are binned according to their overall score on the exam and the log loss is calculated for each model on a test question. Each model follows a common shape, with best performance for high and low performing students which are strongly predictable. Students with scores in the middle are hardest to predict, emphasising how overall ability is the main driver of score prediction.}
    \label{fig:perf_by_cohort}
\end{SCfigure}

Figure \ref{fig:perf_by_cohort} shows how each model performs for students with different overall scores. We define student ability according to their mean score in that exam, excluding a test question. We iterate over all questions and record the log loss for each model on each student in the test dataset. All models perform best for high or low performing students, re-emphasising how overall ability is the primary determinant of model predictions. Students of intermediate ability are much harder to predict, and have a higher log loss. If students had discrete performance profiles that were independent of overall ability, this shape would not be observed.  Additionally, if such a model was deployed to suggest appropriately challenging questions for students, this grade-dependent performance would need to be accounted for to ensure that students in the middle of the performance bands are not sent a large number of less relevant questions to attempt.  

\section{Conclusion}

By comparing several methods, we have demonstrated the applicability of explainable machine learning methods for national-scale prediction of student mathematics results. Similarly to \cite{skill-difficulty}, we find that the dominant factor in predicting student scores was their overall mathematical ability. Our clustering approach has sought to tease out different competency profiles within the data, which could exist due to different locally taught syllabi or students having innate preferences for specific combinations of skills.  \par
We find that a logistic regression model using performance across all questions only narrowly outperforms a single-ability logistic regression, validating previous conclusions that overall ability appears to be a dominant factor in predicting performance. However, the moderate improvement when using every individual question shows that some differences further differences exist between students, albeit at a level that is below the IRT approach in \cite{skill-difficulty}. When building student archetypes using the BMM, there is a strong tendency for these to be dominated by overall ability. However, these clusters can have  varying shapes and profiles that suggest some subgroups could benefit from enhanced personalisation and that these clusters help us to identify such individuals. Furthermore, as overall ability is the main factor in predicting scores, all models perform worse at predicting grades for students with mid-level abilities. This can have important impacts for users of education platforms using such models, potentially leading to students having differing qualities of feedback from the system. \par
 Our work offers a national scale evaluation of using machine learning to understand student mathematical ability in the United Kingdom, offering a new baseline measure. Performance using simpler, explainable models is competitive with deep learning approaches published elsewhere and provides an important benchmark for educational machine learning. If ability is the leading factor, less explainable models may be highly undesirable as they could introduce biases into the predictions and reduce accountability to the students impacted.  Further work is required to explore how personalisation can improve the development of students.


\begin{credits}
\subsubsection{\ackname}This project was funded through the Cheshire and Warrington 4.0 scheme and was completed at the STFC Hartree Centre, which provides funding for small and medium sized enterprises in the Cheshire and Warrington region \href{https://candw4.uk/about-us/}{https://candw4.uk/about-us/}. Additionally, the Scafell Pike HPC facility provided computational resources for this project.  Plots in this manuscript utilised the \texttt{SciencePlots} package in Python \citep{SciencePlots}. We thank Isabel Ainsworth for project management and Richard Harding for assistance with funding applications.

\subsubsection{\discintname}
The authors have no competing interests to declare that are
relevant to the content of this article. 
Conceptualization, Methodology, Software, Validation, Formal analysis, Investigation, Resources, Data Curation, Writing - Original Draft, Writing - Review \& Editing, Visualization, Supervision, Project administration, Funding acquisition
\subsubsection{CRediT statement}
Benjamin Mawdsley: \textit{Software, Methodology, Investigation, Data Curation, Writing - Original Draft, Writing - Review \& Editing, Visualization, Funding Acquisition, Formal Analysis.} Tom Quilter: \textit{Conceptualization, Methodology, Software, Data Curation, Draft, Writing - Review \& Editing, Supervision, Funding acquisition, Formal Analysis.} Richard Turner: \textit{Conceptualization, Methodology, Supervision, Formal Analysis}. Sarah Jackson: \textit{Software, Investigation, Formal Analysis}. Paul Edwards \textit{Software, Investigation, Formal Analysis}
\end{credits}
%
%
%
\bibliographystyle{splncs04}
\bibliography{bibliography}
%





\newpage
\appendix
\section{Further data exploration}
\label{app:exam_desc}

\begin{table}
    \centering
    \begin{tabular}{|c|c|}
    \hline
       Exam number  & Total students after cleaning   \\
       \hline
       Exam 0	 & 32102 \\
Exam 1	& 27770 \\
Exam 2	& 23340 \\
Exam 3	 & 22545 \\
Exam 4	& 18752\\
Exam 5	 &  18313\\ 
Exam 6	 & 19826\\ 
Exam 7	 & 16888\\ 
Exam 8	 & 35365 \\ 
Exam 9	& 30708  \\ 
Exam 10	 & 26447 \\ 
Exam 11	 & 22173 \\ 
Exam 12   & 19559 \\
\hline
\hline
\textbf{\textit{Mean}} & 24137 \\
\hline
    \end{tabular}
    \caption{The number of students completing each exam in our data. As exams are chosen by the local teacher, different numbers of students sit each exam. Students sit multiple exams in the data - see Figure \ref{fig:sit_both}. We clean the data to remove issues such as scores above 100\%.}
    \label{tab:exam_counts}
\end{table}

Table \ref{tab:exam_counts} shows how many students completed each exam. Pupils sit different combinations of these exams based on the choices of their teacher. The exam with most students was Exam 8, whereas the one with the fewest was 7, but there is only approximately a factor of 2 different between them and the exams have a similar order of magnitude of samples. \par

\begin{SCfigure}[1.0][!ht]
    \centering
    \includegraphics[width=0.5\columnwidth]{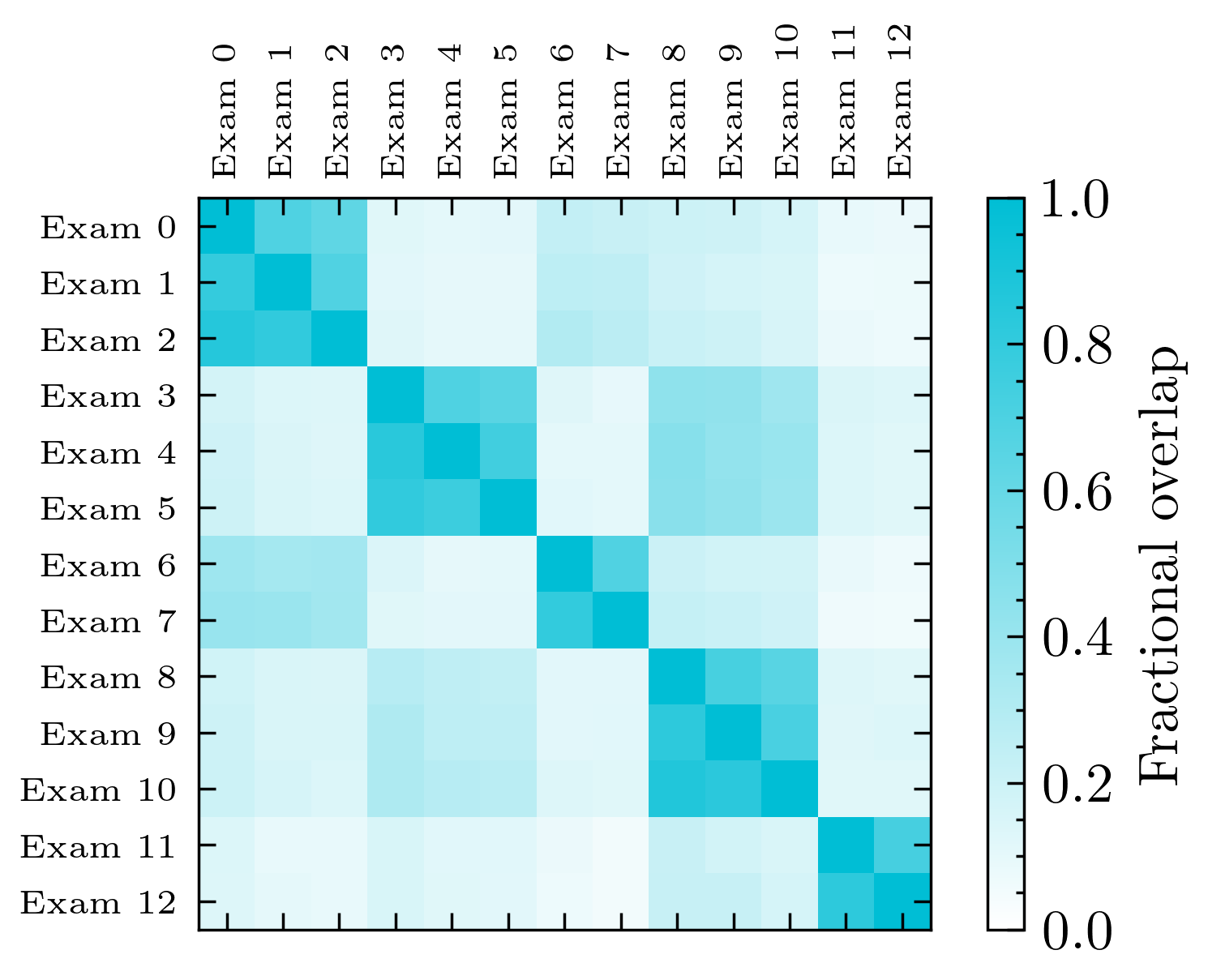}
    \caption{Visualising the degree of overlap in the exams sat by the students in our dataset. Most students have a high degree of overlap between a few exams in our dataset, but a small amount of overlap with other exams outside of that core range. }
    \label{fig:sit_both}
\end{SCfigure}

Figure \ref{fig:sit_both} gives an indication of the extent of overlap between the different exams, with the colours in the correlation matrix denoting the number of students that have sat both exams. As the mock exams are selected locally within a school, there is no expectation that all exams are sat by all students in the dataset. A clear preference can be seen where some exams are more likely to be sat together. For example, Exams 8,9 and 10 are all very likely to be shared by a student. 

\begin{figure*}
    \centering
    \includegraphics[width=\textwidth]{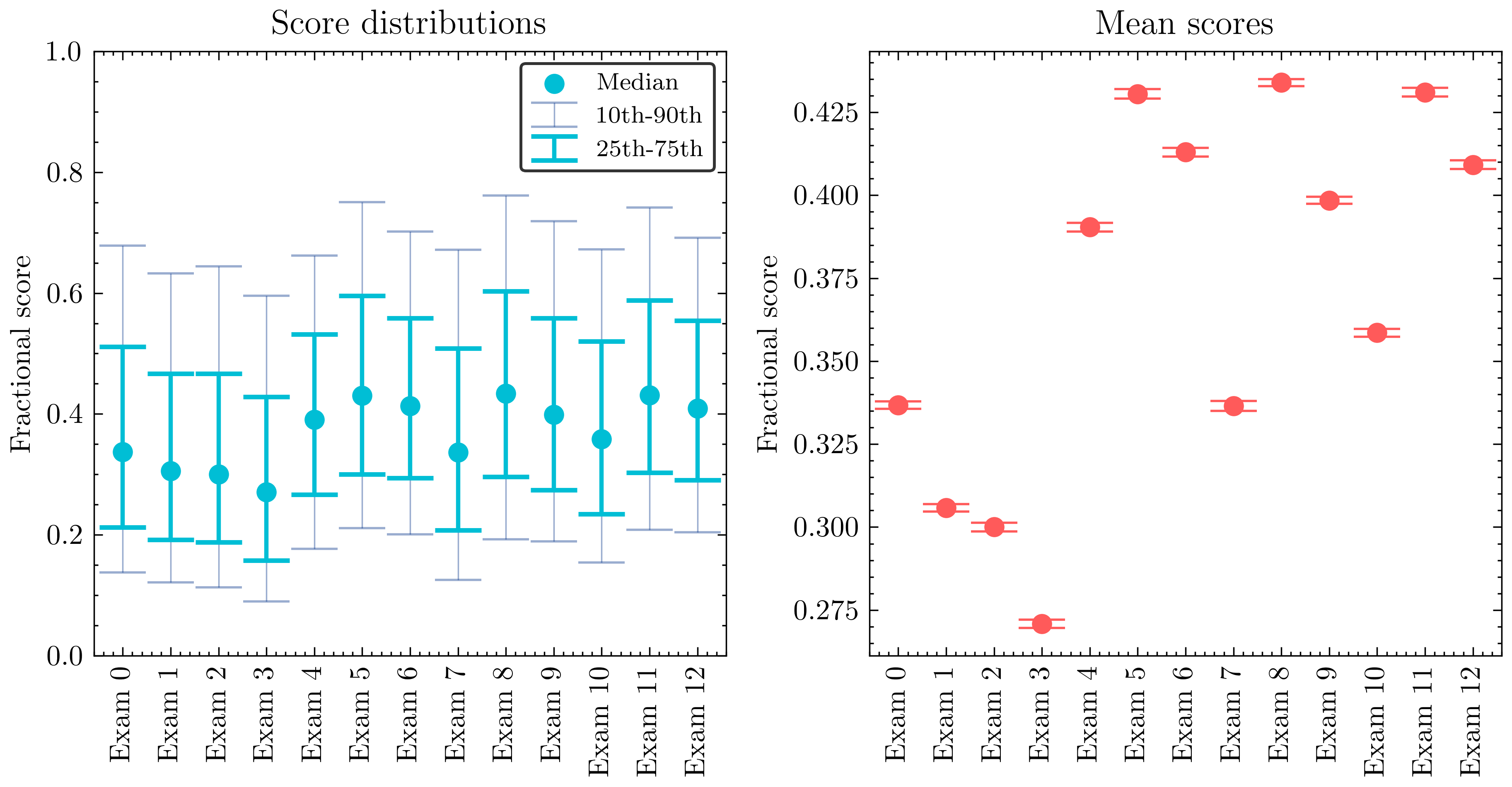}
    \caption{The overall fractional scores of students across the exams, calculated when each question is treated as a binary pass or fail. Left: the average score in each exam with error bars showing the percentile ranges of scores in that exam. Right: the mean score in each exam, with error bars denoting the error on the mean ($\sigma / \sqrt{N}$). From the left, exams broadly cover the same range of scores, but from the right it can be seen that small differences in performance (or exam difficulty) are present.  }
    \label{fig:exam_score_dist_vsbinary}
\end{figure*}

The left panel of Figure \ref{fig:exam_score_dist_vsbinary} shows the mean score in each of the exams used in this analysis, with some percentile ranges of that distribution overlaid to show the distribution of scores. The panel on the right shows the mean score in each exam, with error bars now denoting the standard error on this mean. Whilst broadly comparable, small variations in mean performance across exams could be due to statistical variation, slight fluctuations in difficulty, or the different populations sitting each exam (see Figure \ref{fig:exam_score_dist_vsbinary} in Appendix \ref{app:exam_desc} for a deeper exploration.)  Due to the large number of students in each exam, this mean value is tightly constrained. Exam 3 has a much lower mean score than Exam 11, for example, suggesting significantly different exam difficulty and/or student competencies in those exams. Exams 0,1 and 2 are sat by similar cohorts in Figure \ref{fig:sit_both}, meaning that Exam 0 appears to be easier than Exams 1 and 2, as there is a lower typical performance in the latter two.

\begin{figure*}[!h]
    \centering
    \includegraphics[width=\linewidth]{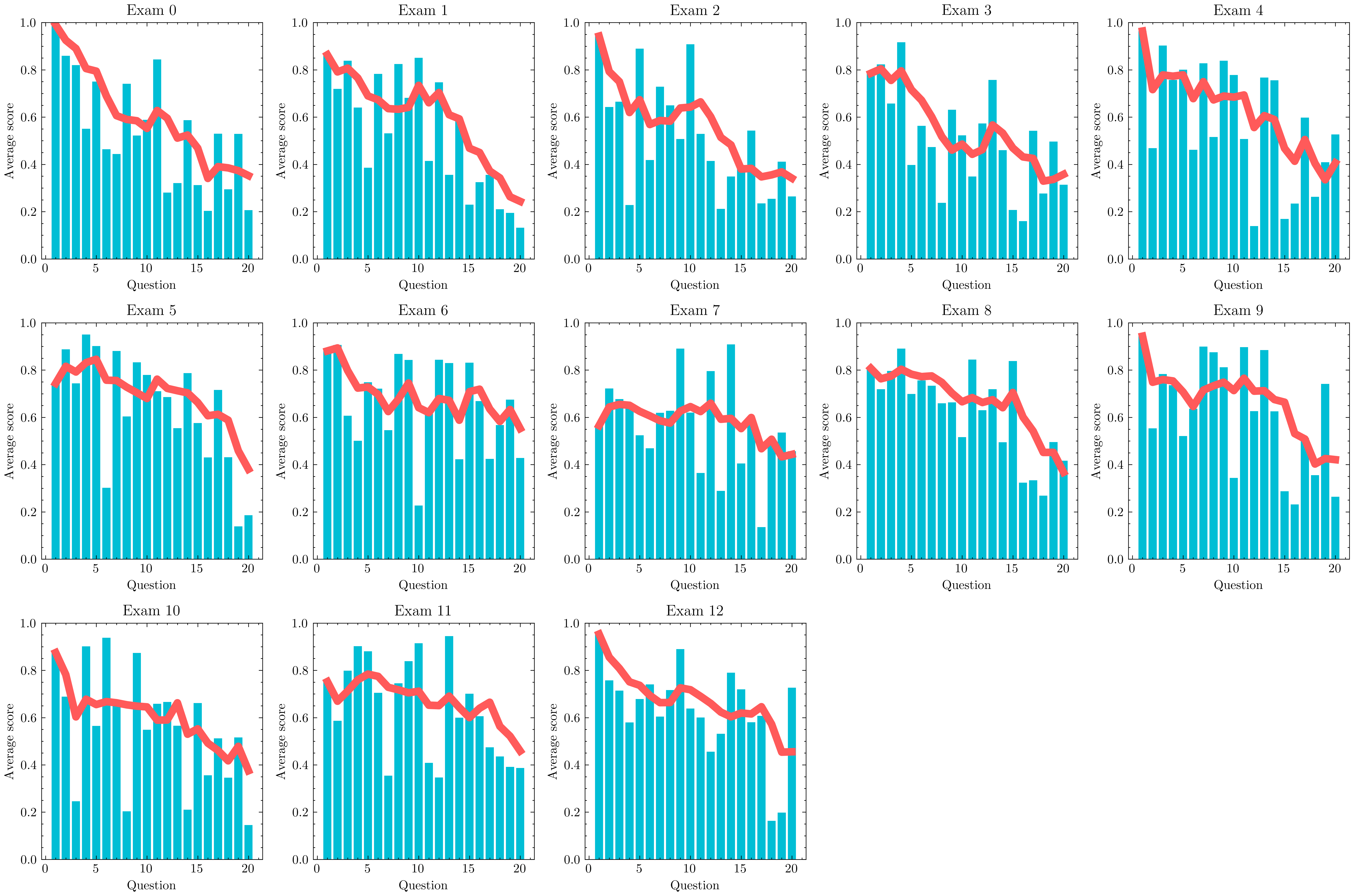}
    \caption{The average scores in each question across all exams in our dataset. The bars show the average score on that specific question, whilst the red line shows the rolling average, considering that question alongside the three preceding questions and weighting them all equally. More difficult questions are usually placed at the end of the exam and a drop off in typical score can be seen. We include questions 1-20 only as these should be completed by all participants, whereas questions 21-25 have optional components. }
    \label{fig:scores_byq}
\end{figure*}

For added granularity, Figure \ref{fig:scores_byq} shows how the average score across a \textit{question} varies through each of the exams used in our dataset. Lower scores at latter questions can be explained by two factors: harder questions are typically at the end of the exam, and students that struggle to solve earlier questions are less likely to complete the latter questions. This trend is visible across all exams, with the severity of the decline varying in each exam. For example, Exam 0 has a very steep decline in average score in the latter half of the exam, whereas Exam 5 has a much flatter profile. Exam 5 appears to have its most difficult question midway through, which is unusual within this dataset. Within each exam, there is significant variation in the typical grade in each question, suggesting that different factors are in play for performance in each question. \par 

\section{Exploring inter-cluster correlations}
\label{app:intercluster_corr}
\begin{figure*}[!ht]
    \centering
    \includegraphics[width = \linewidth]{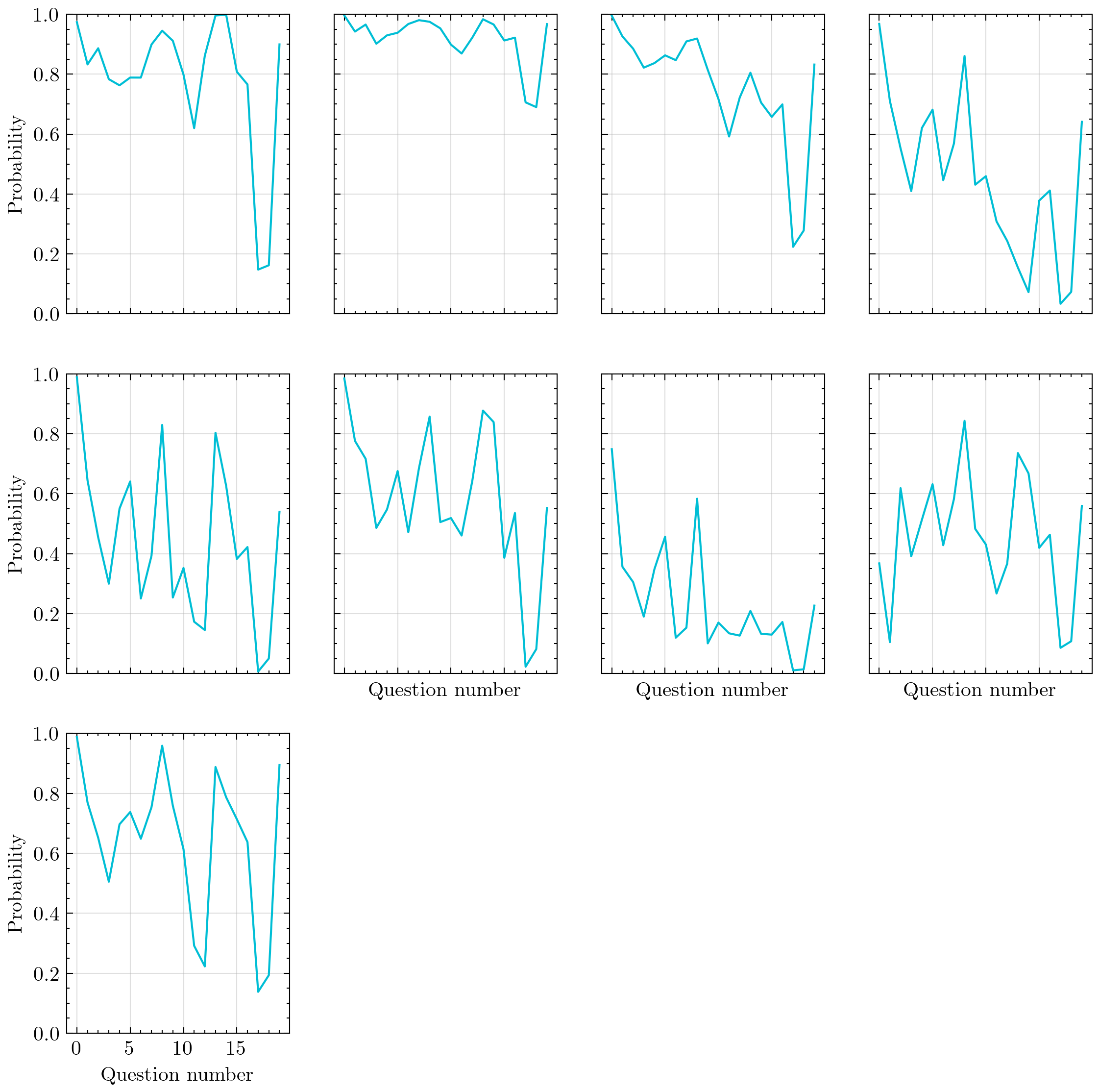}
    \caption{ A series of clusters with a high linear correlation. Overall, most clusters have a similar shape to their profiles, causing a high linear correlation, and showing that performance on each question is generally dominated by the students' overall mathematical ability.}
    \label{fig:bmm_mostcorr}
\end{figure*}

 Looking at individual cluster fits can further emphasise the linear correlations suggested by Figure \ref{fig:intercluster_linear}. The vast majority are positive correlated (\ref{fig:bmm_mostcorr}), and an example set of highly correlated clusters can be seen in Figure \ref{fig:bmm_mostcorr}. Each panel shows a separate cluster output by the expectation maximisation routine, with the x axis being each question in the exam and the y axis being the probability of getting that question right. The overall profile of each line plot in this set is broadly similar - each has a peak at question 5, 8, and 13, but the overall amplitude of it scales different in each cluster. Most clusters all perform badly on questions 17 and 18, with only the highest performing cluster number 2 reaching a high probability of getting that question right (albeit this question still being the worst for this cluster). Interestingly, there does seem to be a slight difference between clusters 1, 2 and 3; all perform quite well on the exam overall, but cluster 1 performs best on questions 13 and 14 but worst on 17 and 18. This highlights some of the small variations in performance that these clusters can capture, above a flat single-ability model.  \par

\begin{figure*}[!ht]
    \centering
    \includegraphics[width=\linewidth]{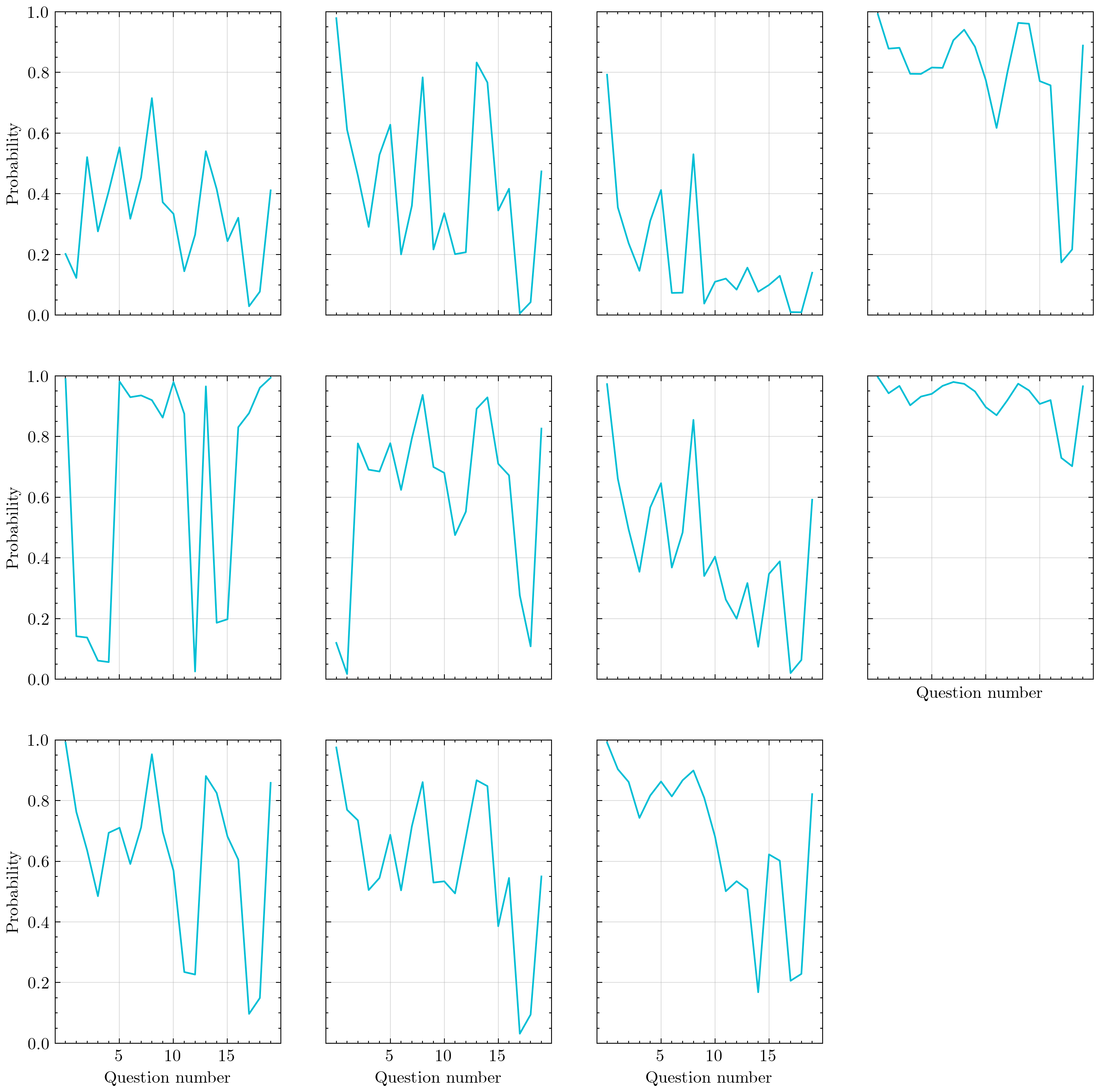}
    \caption{ A series of clusters with a low linear correlation. This low level of correlation is relatively rare in the dataset, but the shape of the clusters visually differ much more widely than those shown in Figure \ref{fig:bmm_mostcorr}.}
    \label{fig:bmm_leastcorr}
\end{figure*}

 At the opposite end of the scale, Figure \ref{fig:bmm_leastcorr} examines a set of clusters with a very low linear correlation to see how the approach can find widely differing skill profiles. In this exam, there are two different populations of high performers (clusters 4 and 8) that perform well across most questions, but with cluster 8 scoring much more highly on questions 17 and 18. Most other clusters look different to each other, with a few common peaks around questions 5 and 7 and worse performance at question 11. Cluster 5 has a very unusual profile, with some of the easier early questions having a very low probability of correctness before having a high success rate in 5-12 questions; different teaching schedules could cover topics in different orders, meaning these highly capable students may not have met the topics in questions 2-5 in this exam, compared to most of their peers.   

\end{document}